\documentclass[reprint,aps,pra,showkeys,showpacs]{revtex4-1}  
\usepackage{epsfig}
\usepackage{xcolor}   

\begin{document}

\title{The variational approximation for two-dimensional quantum droplets}

\author{Sherzod R. Otajonov}
\author{Eduard N. Tsoy}
\author{Fatkhulla Kh. Abdullaev}

\affiliation{Physical-Technical Institute of the Uzbek Academy of Sciences,\\
Chingiz Aytmatov str. 2-B, Tashkent, 100084, Uzbekistan}

\date{\today}

\begin{abstract}
  The dynamics of a two-dimensional Bose-Einstein condensate in a presence of quantum
fluctuations is studied. The properties of localized density distributions,  quantum droplets
(QDs), are analyzed by means of the variational approach. It is demonstrated that the
super-Gaussian  function gives
a good approximation for profiles of fundamental QDs and droplets with non-zero vorticity.
The dynamical equations for parameters of QDs are obtained. Fixed points of these equations 
determine the parameters of stationary QDs. The period of
small oscillations of QDs near the stationary state is estimated.  It is obtained that
periodic modulations of the strength of quantum fluctuations can actuate different
processes, including resonance oscillations of the QD parameters, an emission of waves
and a splitting of QDs into smaller droplets.
\end{abstract}

\keywords{Quantum droplets; Bose-Einstein condensate; two-dimensional solitons;
quantum fluctuations}

\maketitle

\section{Introduction}
\label{sec:intro}

   The quantum pressure and the two-body interaction (2BI) are the basic effects that
determine the mean-field dynamics of Bose-Einstein condensates (BECs)~\cite{Peth08}. A
balance between these two effects can result in a formation of matter-wave solitons.
Solitons are stable in one-dimensional (1D) systems with attractive cubic nonlinearity, but they
are unstable in higher dimensions. Different methods are suggested to stabilize localized waves
in 2D and 3D BECs. These methods include an account of higher-order nonlinearities, an
application of external traps, an account of the dipolar interaction, and a dynamical
variation of the interaction parameter~\cite{Peth08,Kart19}.

   Recently, it was suggested that quantum fluctuations (QFs) can stabilize localized waves
in BECs~\cite{Petr15,Petr16}. Moreover, it was shown that QFs are responsible for a
formation of quantum droplets (QDs), which are clusters of an ultra-dilute liquid. Quantum
fluctuations are effects beyond the mean-field description of BECs. These effects are
described by the Lee-Huang-Yang (LHY) correction term~\cite{LHY} in the BEC
Hamiltonian. Usually, the influence of QFs is small comparing with the 2BI. A decrease of
the 2BI parameter via the Feschbach resonance does not help to reveal QFs, since the
parameter of the LHY term diminishes as well. However, in BEC mixtures, it is possible to
tune the parameters of intra-species and inter-species interactions in such a way that the
residual 2BI is comparable with QFs~\cite{Petr15,Petr16}. Indeed, a formation of QDs in binary
BECs was observed experimentally~\cite{Cabr18,Seme18}.

   A dipolar BEC is another example of atomic systems, where the effect of QFs can be made
pronounced. The strength of the dipolar interaction can be tuned independently of the LHY
parameter. Then, the dipolar interaction can almost balance the 2BI, making QFs
appreciable. An ultra-dilute liquid and QDs were also observed experimentally in dipolar
BECs~\cite{Ferr16}.

   In the present paper, we study QDs in a binary 2D BEC.
Basically, QDs are solitons induced by quantum fluctuations. We consider such a situation,
when the dynamics of the BEC components is described by a single equation. This situation is
possible in the symmetric case, when the component densities, as well as the coupling constants, 
are close to each other, see Ref.~\cite{Petr16} and Sec.~\ref{sec:model}. Such a condition was 
used in other studies~\cite{Petr16,Li18,Kart19a} as well. We apply the variational approximation (VA),
and demonstrate that the super-Gaussian function describes well QDs in a wide range of the
system parameters. We obtain explicit relations that define stationary and dynamical
parameters of two-dimensional QDs. A response of QDs to the periodic modulation in time of
the LHY parameter is analyzed as well.

\section{Results}
\label{sec:res}

\subsection{The model and the variational approximation}
\label{sec:model}

   We consider a binary 2D BEC with the repulsive intra-species interaction and attractive 
inter-species interaction. The dynamics of the BEC in a presence of quantum fluctuations is
governed by the following equations~\cite{Petr16}
\begin{eqnarray}
&&   i \hbar \frac{\partial \phi_j}{\partial T} =  - \frac{\hbar^2}{2m} \nabla^2 \phi_j + 
  \left[ (-1)^{j-1} g_j^{1/2} \times  \right.
\nonumber \\   
&&  \left.  (g_1^{1/2} |\phi_1|^2 - g_2^{1/2} |\phi_2|^2)  +
     \frac{g_j P}{4 \pi \eta} \log \frac{P e}{\eta \Delta} \right] \phi_j ,
\label{set}
\end{eqnarray}
where $\phi_j$ are the component wave functions, 
$\nabla^{2} = \partial_X^2 + \partial_Y^2$, $T$ is time,
$
g_j = {4 \pi \eta \over \log[4 e^{-2C} /(a_j^2 \Delta)]}
$
are the modified coupling constants, $P= g_1 |\phi_1|^2 + g_2 |\phi_2|^2$, 
$
\Delta = {4 e^{-2C} \over |a_x| \sqrt{a_1 a_2}} \exp\{ {- \log^2(a_2/a_1) \over 2 \log[a_x^2 / (a_1 a_2)]} \}
$,
$\eta = \hbar^2/m_a$, $j = 1$ and $2$, $m_a$ is the atom mass,
$a_1$ and $a_2$ ($a_x$) are the 2D intra-species (inter-species) scattering
lengths~\cite{Petr16}, and $C \approx 0.5772$ is the Euler constant.

   In the symmetric case, $\phi_1 = \phi_2 = \phi$ and $g_1 = g_2 = g$, Eqs.~(\ref{set})
are reduced to the single Gross-Pitaevskii equation~\cite{Petr16,Li18} in the dimensionless form
\begin{equation}
   i \partial_{t} \Psi+ {1 \over 2} \nabla ^{2} \Psi + \gamma |\Psi |^{2} \Psi + \delta |\Psi|^{2}
   \log (|\Psi|^{2}) \Psi =0,
\label{gpe}
\end{equation}
where $\Psi= \phi / \sqrt{n_0}$, $\nabla^{2} = \partial_y^2 + \partial_y^2$, 
$x = X/X_{\mathrm{sc}}$, $y = Y/Y_{\mathrm{sc}}$, $t = T/T_{\mathrm{sc}}$, 
$\delta= -g^2 n_0 m/ (\pi \hbar^3 \omega_0)$ characterizes the strength of QFs, 
$\gamma = -\delta /2$, and $n_0 = \Delta / (2 g e^{3/2})$ is the equilibrium density of a 
component~\cite{Petr16}. The time scale  and the spatial scale are defined in Eq.~(\ref{gpe})
as $T_{\mathrm{sc}} = \omega_0^{-1}$, and 
$X_{\mathrm{sc}} = Y_{\mathrm{sc}} = [\hbar / (m \omega_0)]^{1/2}$,
respectively, where $\omega_0$ is the characteristic frequency of the external trapping
potential~\cite{Peth08}. When $\gamma > 0$ and $\delta = 0$, there are no stable
localized solutions in the system. For these values, solitons either decay or collapse,
depending on initial conditions~\cite{Garm64}. Quantum fluctuations, described by 
the LHY term with $\delta < 0$, arrest the collapse~\cite{Petr15,Petr16}.

   There are certain indications that the symmetric state is stable. A deviation from the symmetric 
state adds a non-negative term to the energy~\cite{Petr16}. Also, numerical simulations for the 
non-symmetric case in Ref.~\cite{Li18} show results consistent with the symmetric case.  
Below, we assume that small perturbations of the symmetric state do not change substantially 
the dynamics.

   The Lagrangian density of Eq.~(\ref{gpe}) is defined as follows
\begin{eqnarray}
    \mathcal{L} = && {i \over 2} (\Psi^{*} \partial_{t} \Psi  - \Psi \partial_{t} \Psi^{*})
     - {1 \over 2} \left( |\partial_{x} \Psi|^{2} + |\partial_{y} \Psi|^{2} \right)
\nonumber \\
     && +{\gamma \over 2} |\Psi|^{4} + {\delta \over 2} |\Psi|^{4}
       \log \left({|\Psi|^{2} \over \sqrt{e} }  \right).
\label{lagrdens}
\end{eqnarray}
By using transformation $\Psi \exp[\gamma /(2\delta)] \to \bar{\Psi}$ and
$\delta \exp(-\gamma / \delta) \to \bar{\delta} $, we can eliminate terms proportional to
$\gamma$ in Eqs.~(\ref{gpe}) and~(\ref{lagrdens}), even when $\gamma$ and $\delta$ are independent. 
Therefore, below we take $\gamma = 0$, however, results obtained are valid also for the general case.
The energy density $\mathcal{E}$  of a BEC is related to $\mathcal{L}$ as
\begin{equation}
  \mathcal{E} = {i \over 2} (\Psi^{*} \partial_{t} \Psi  - \Psi \partial_{t} \Psi^{*}) - \mathcal{L}.
\label{en}
\end{equation}

    Function $V(x,t) \equiv -\delta |\Psi|^2 \log|\Psi|^2$ in the last term of Eq.~(\ref{gpe}) can be
considered as an effective self-induced potential for field $\Psi$. When $\delta < 0$, a pulse-shaped
distribution of the BEC density with $|\Psi|^2 < 1$ results in an attractive potential $V(x,t)$. 
This attractive force due to quantum fluctuations can balance the quantum pressure, described 
by the second term in Eq.~(\ref{gpe}), so that a formation of localized waves
(vortices) is possible. 

   There are no known solutions of Eq.~(\ref{gpe}). In order to characterize QDs, we employ the 
following super-Gaussian trial function:
\begin{equation}
  \Psi(r, \theta, t)= A r^{S} \exp\left[ -{1 \over 2} \left({r \over w} \right)^{2m} +
    i (b r^{2} \pm S \theta + \varphi) \right],
\label{ansatz}
\end{equation}
where $A(t)$, $w(t)$, $b(t)$, and $\varphi(t)$ are the variational parameters, denoting the
amplitude parameter, width, chirp, and the initial phase, respectively. Parameter $m$
defines the profile shape. When $m \approx 1$ and $S = 0$, the profile has a bell shape, while when
 $m \gg 1$ ($0 < m \ll 1$), the profile tends to a flat-top (cusp) shape. The non-negative integer
parameter $S\ge 0$ is the topological charge (vorticity) of a QD. The fundamental
(zero-vorticity) QD has $S = 0$, while vortex QDs have $S > 0$. The plus (minus) sign in a front 
of $S$ corresponds to a vortex (anti-vortex). Parameter $A$ is related to the
maximum $\bar{A}^2$ of the QD density as the following
\begin{equation}
  \bar{A} = A\, (M S)^{M S/2} w^{S} \exp(-MS/2),
\label{ampl}
\end{equation}
where $M = 1/m$. When $S = 0$, the QD size is specified by parameter $w$, such that
the full width at half maximum, $w_\mathrm{FWHM}$ is found as
$w_\mathrm{FWHM}= 2 (\log 2)^{M/2} w$. When $S > 0$, the
QD size can be defined as a position of the density maximum, $r_{\mathrm{max}} = (M S)^{M/2}
w$.

   When $A(t) = \mathrm{const}$ and $w(t) = \mathrm{const}$, Eq.~(\ref{ansatz})
gives an approximation of a stationary solution $\Psi(r,\theta, t) = u(r, \theta) \exp(-i\mu t)$ of Eq.~(\ref{gpe}),
where $u(r, \theta)$ is a profile function, and $\mu$ is the chemical potential, see Eq.~(\ref{mu}).
We mention that the super-Gaussian function describes well (with accuracy 1-5\%) 1D
quantum droplets~\cite{Otaj19}. As it is shown below, this function gives also a reasonable
approximation for profiles and parameters of 2D QDs.

   Norm $N$ is the conserved quantity of Eq.~(\ref{gpe}), and it is proportional to the
number of atoms in a BEC cloud. In terms of the trial function, $N$ is written as
\begin{equation}
     N = \int \limits_{\ -\infty}^{\ \ \infty} \! \! \! \! \! \int
     {|\Psi|^{2} dx\, dy} =\pi M A^{2} w^{2(S+1)} \Gamma(M(S+1)) .
\label{norm}
\end{equation}

   Substituting the trial function into Eq.~(\ref{lagrdens}), and integrating over the
spatial variable, we get the averaged Lagrangian $L = \int \! \! \int {\mathcal{L} dx dy}$
\begin{eqnarray}
 {L \over N}= -\varphi' - {\Gamma(M(S+2)) \over \Gamma(M(S+1))}\, w^{2} (2b^{2} +b') - G,
\label{lagr}
\end{eqnarray}
where the prime denotes the time derivative, and
\begin{eqnarray}
  G&&(w, N, M) = { \Gamma(M S+2) \over 2 M^{2} \Gamma(M(S+1)) w^{2} }
\nonumber \\
  && + {N \Gamma(M(2S+1)) \over 2^{2+M(2S+1)} \pi M \Gamma^{2}(M(S+1)) w^{2}}
\nonumber \\
  && \times\, \delta \bigg\{ 1 + M  + 2 M S \left[ 1-\psi(M(2S+1)) \right]  \bigg.
\nonumber \\
  && \left. - 2 \log \left[ {N \over 2^{MS} \pi M w^{2} \Gamma(M(S+1))} \right] \right\} .
\label{G}
\end{eqnarray}
Here $\Gamma(z)$ is the Gamma function, and $\psi(z)= d \ln \Gamma(z) / dz$ is the digamma
function. The Euler-Lagrangian equations for $L$ result in the following set of equations for the
QD parameters:
\begin{eqnarray}
  b' &=& -2{{b}^{2}}-{\Gamma(M(S+1)) \over 2\Gamma(M(S+2))} \, {1 \over w}
    {\partial G \over \partial w} \equiv f_{b},
\nonumber \\
  w' &=& 2bw \equiv f_{w},
\nonumber \\
  f_{m} &\equiv& \frac{\partial L}{\partial m} = 0 .
\label{pars}
\end{eqnarray}
The parameters of a stationary QD are found from $b = 0$, $f_b(w, N, m) = 0$, and
$f_m(w, N, m) = 0$. Therefore, similarly to Ref.~\cite{Otaj19}, we state that for a given
set of the system parameters, index $m$ takes the value that corresponds to a stationary
state.

\begin{figure}[htbp]
  \centerline{ \includegraphics[width=6cm]{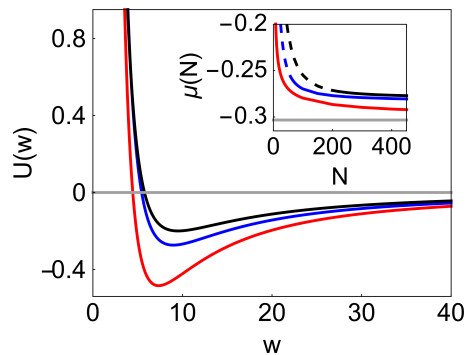}}
\caption{(color online) The shape of effective potentials for different parameters of $(S, N, m)$.
The bottom (red), middle (blue) and top (black) lines are for $(0, 100, 2.537)$, $(1, 200, 1.971)$,
and $(2, 200, 2.073)$, respectively. The inset shows the chemical potential $\mu$ as a
function of norm $N$, see Eq.~(\ref{mu}). The order (color) and parameters of the curves are the same
as in the main plot. Dashed lines for $S = 1$ and $2$ correspond to numerically
unstable regions. The gray line in the inset represents the Thomas-Fermi limit.}
\label{fig:u_mu}
\end{figure}

   The first two equations of Eqs.~(\ref{pars}) can further be combined into a single equation
for the QD width:
\begin{equation}
  w'' = -{\Gamma(M(S+1)) \over \Gamma(M(S+2))}\,
    \frac{\partial G}{\partial w} \equiv -{\partial U(w) \over \partial w}.
\label{wtt}
\end{equation}
Therefore, the dynamics of the QD parameters is reduced to the dynamics of an effective
particle with coordinate $w$ in a potential $U(w)$, see also Ref.~\cite{Otaj19}:
\begin{equation}
  U(w) = { \Gamma(M(S+1)) \over \Gamma(M(S+2)) }\, G.
\label{pot}
\end{equation}
The potentials for different values of $(S, N, m)$ are plotted in Fig.~\ref{fig:u_mu}. 
The minimum of the potential [or a zero of $f_b(w, N, m) = 0$] corresponds to width
$w_{s}$ of a stationary quantum droplet:
\begin{equation}
   w_s = {\sqrt{N} \over \sqrt{2^{MS} \pi M \Gamma(M(S+1))}}\, e^c,
\label{ws}
\end{equation}
where
\begin{eqnarray}
   c &=& {1 \over 4} \bigg[ 1 - M(2S+1) + 2 M S \psi(M(2S+1)) - \bigg.
\nonumber   \\
   && \left. 2^{1+M(2S+1)}  {\pi \Gamma(MS+2)\Gamma(M(S+1)) \over \delta M N \Gamma(M(2S+1))}
   \right] .
\end{eqnarray}

   For given $S$ and $N$, the width $w_s$ depends only on $m$, which is determined from
$f_m(w_s, N, m)=0$. Then, the the amplitude parameter $A_s$ of a stationary droplet is obtained
from Eq.~(\ref{norm}), while the chemical potential $\mu \equiv -\varphi'$
is found as
\begin{equation}
  \mu= \left.{\partial E_{s} \over \partial N}\right|_{w=w_{s}} =
    \left. {\partial (NG) \over \partial N}\right|_{w=w_{s}}.
\label{mu}
\end{equation}
where $E_{s} =\int \! \! \int {\mathcal{E} dx dy} = NG$ is the energy of the stationary QD.
Thus, the VA provides equations for calculation of main parameters of stationary QDs.
The inset of Fig.~\ref{fig:u_mu} shows the dependencies of the chemical potential $\mu$ on
norm $N$ for the values of $S=0, 1$, and $2$. The chemical potential approaches a constant
value at large $N$. This is typical for flat-top solitons. For a reference, the
Thomas-Fermi limit $\mu_\mathrm{TF}=-1/(2 \sqrt{e})$~\cite{Li18}, valid for large $N$, is also
shown in the inset of Fig.~\ref{fig:u_mu}. Analysis of other parameters of stationary QDs
are presented in Sec.~\ref{sec:num}. We provide also approximate equations for $m$ so that
the QD parameters can be calculated directly.

  The frequency of small oscillations near the stationary width is obtained by using
Eq.~(\ref{pot})
\begin{equation}
  \Omega^{2}_{0} = \left. {\partial^{2} U(w) \over \partial w^{2} } \right|_{w=w_{s}}
    ={\Gamma(M(S+1)) \over \Gamma(M(S+2))} \left. {\partial^{2} G \over \partial w^{2} } \right|_{w=w_{s}}.
\nonumber \\
\label{freq}
\end{equation}
The frequency $\Omega_0$ defines the frequency of the Goldstone mode of a QD, oscillating near
the stationary state. This parameter can be used in experiments to estimate the strength of QFs.

   Quantum fluctuations are described by the LHY term with $\delta < 0$, however the VA
allows us to analyze the opposite case as well. When $\delta > 0$, the potential $U(w)$ tends to
$-\infty$ at $w \to 0$, to $0$ at $w \to \infty$, and has a single maximum at $w = w_s$,
which corresponds to an unstable stationary solution. This value is small, $w_s \lesssim
0.5$ for $\delta = 1$ and $N = [1,1000]$. Moreover, for $S = 0$ and $N \gtrsim 6.3$ the value of
$\mu$ becomes positive, and $m < 0.5$. Value $w_s$ is a threshold that separates different types
of the dynamics. For given $N$, if the initial width is larger (smaller) than the
threshold, then the soliton decays dispersively (collapses). In numerical simulations, the
threshold value is $\sim (1.05 \textrm{--} 1.1)\, w_s$. A presence of a collapse has a
simple physical explanation. When $\delta > 0$ and $|\Psi|^2 > 1$, the last term in Eq.~(\ref{gpe}) 
corresponds to self-attraction. Since the order of nonlinearity is greater than cubic, the attraction
cannot be balanced by dispersion, resulting in a collapse in the system.

\subsection{Numerical simulations}
\label{sec:num}

\begin{figure}[htbp]
\centerline{ \includegraphics[width=6cm]{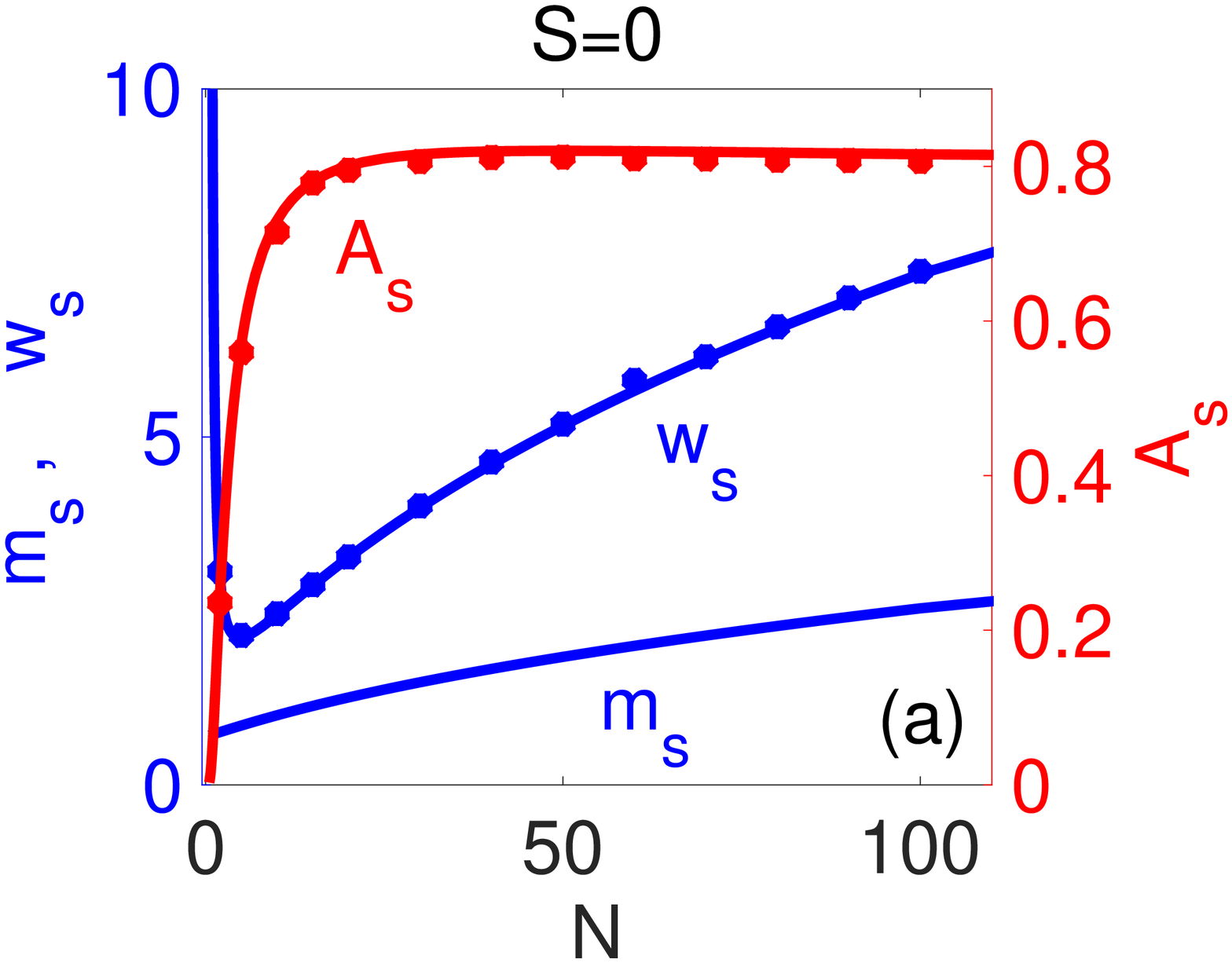} } 
\centerline{ \includegraphics[width=6cm]{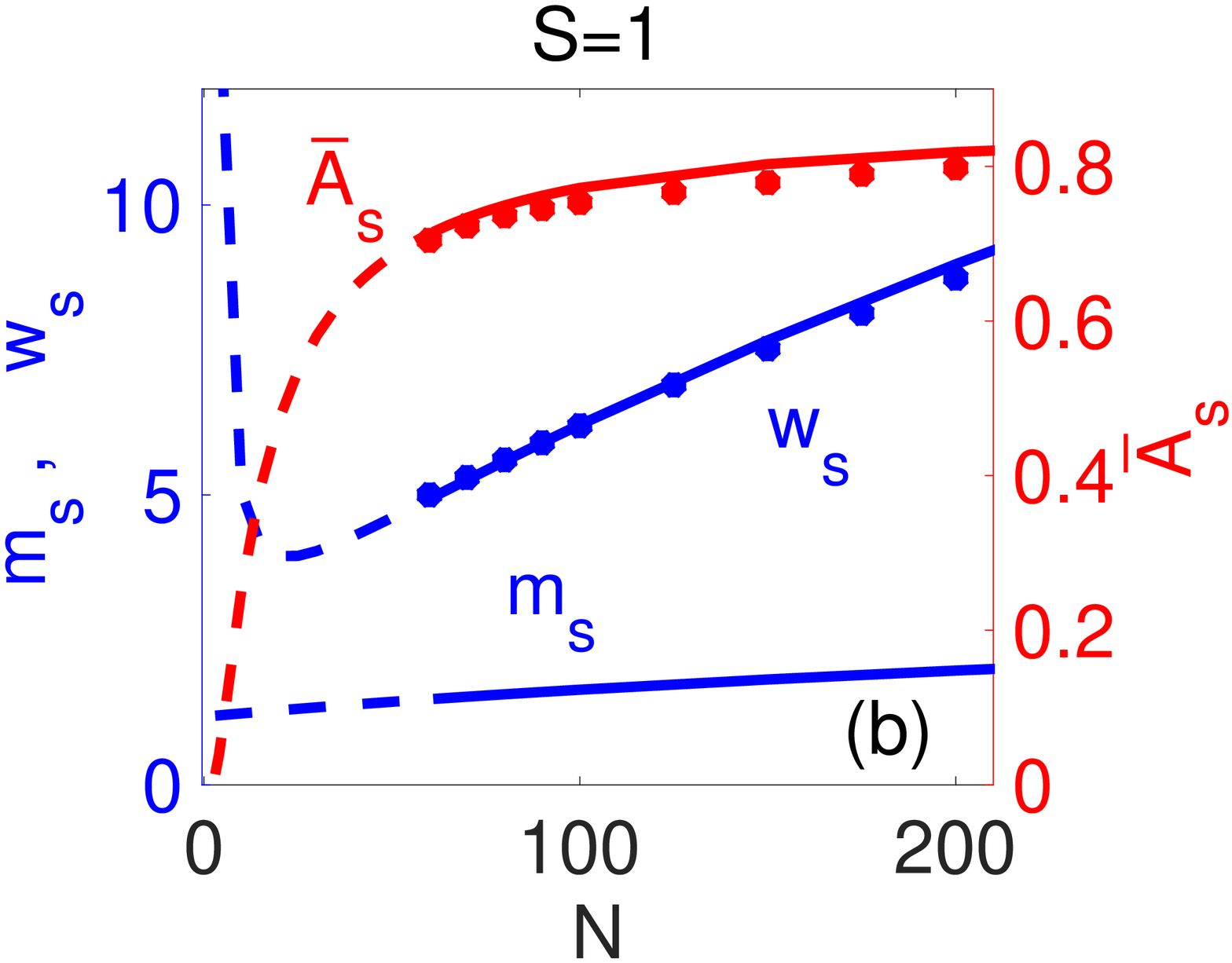} }
\caption{(color online) (a) Parameters of stationary QDs, found from the VA (lines) and
from numerical simulations (points) of Eq.~(\ref{gpe}) for $S = 0$. The left axis is for
QD width $w_s$ and $m_s$, while the right axis is for parameter $A_s$.
(b) The same as in panel (a) but for $S = 1$. Dashed lines correspond to numerically
unstable regions.}
\label{fig:pars}
\end{figure}

   A comparison of the stationary QD parameters, found from the VA, with those, found from
numerical simulations of Eq.~(\ref{gpe}), is presented in Fig.~\ref{fig:pars}. In numerical
simulations of Eq.~(\ref{gpe}), we take an initial condition in form~(\ref{ansatz}) with
parameters $A_s$, $w_s$, and $m_s$, found from the VA.  Then Eq.~(\ref{gpe}) is integrated
by the split-step Fourier (SSF) method with 512$\times$512 discrete points and the spatial region of
size $d\times d$, where $d \sim 50$--$100$, depending on the QD width. Absorbing boundary
conditions are used to prevent reflections of waves, emitted by a QD, from the end points of the 
region.

   An exact stationary solution
has a time-independent spatial distribution of the BEC density. However, we observe small
oscillations on time of the QD parameters near a stationary state. The relative amplitude of oscillations
is less then few percent of the stationary value that shows an acceptable accuracy of the VA.
An additional change of the initial parameters by $\sim 1 \textrm{--} 5\%$ results just in
a corresponding small increase of the amplitude of oscillations, indicating the stability of the stationary
state. The amplitude parameter $A$ is found from the maximum $\bar{A}^2$ of the field density 
and Eq.~(\ref{ampl}). The QD width is found numerically from the following equation
\begin{equation}
  w^{2}= {\Gamma(M(S+1)) \over \Gamma(M(S+2)) N }
    \int \limits_{\ -\infty}^{\ \ \infty} \! \! \! \! \! \int
    { (x^{2}+y^{2}) |\Psi|^{2}  dx dy } .
\end{equation}
We mention that this equation gives acceptable values of the width when deviations
from the stationary profile are small. Stationary values of the QD parameters are found as an average
on time after several initial oscillations.
Figure~\ref{fig:pars} demonstrates that the VA gives a good prediction, and that the super-Gaussian
profile is close to the actual stationary solution of Eq.~(\ref{gpe}).
The behavior described above is valid for sufficiently large $N$. For smaller $N$, stationary states can
be unstable (see below).

   It follows from Fig.~\ref{fig:pars} that the QD amplitude $A_s$ tends to a constant value for
large $N$, while the width increases on $N$. This fact and a gradual increase of $m_s$ mean
that a QD approaches the flat-top shape for large $N$. For practical purpose, we
approximate dependence $m_s(N)$ for $\delta = -1$ as
\begin{eqnarray}
  m_s  &=&  (0.4433 + 0.05906 N)^{0.5047} \quad \mbox{ for } S = 0,
\nonumber \\
  m_s  &=&  (1.519 + 0.02415 N)^{0.3692}  \quad \mbox{ for } S = 1.
\label{m(n)}
\end{eqnarray}
A functional form of $m_s(N) = (a_1 + a_2 N)^{k}$ is taken empirically, and parameters are
found by fitting with values of $m_s$, found numerically from the VA for $0 < N \leq 1000$. 
Using Eqs.~(\ref{m(n)}),
we can find directly the stationary width $w_s$ from Eq.~(\ref{ws}), then $A_s$ from Eq.~(\ref{norm}),
and $\mu$ from Eq.~(\ref{mu}).

\begin{figure}[htbp]
\centerline{ \includegraphics[width=7.2cm]{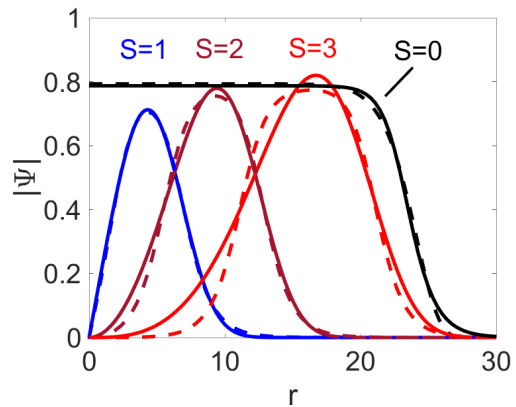}}
\caption{(color online) 
Profiles $|\Psi|$ of QDs, found from the VA (solid lines) and  from the imaginary-time 
method (dashed lines) for $S=0, 1, 2$ and 3, and $N= 1000, 60, 200$ and 510, respectively.
}
\label{fig:prof}
\end{figure}

   In Fig.~\ref{fig:prof}, we present a comparison of stationary QD profiles, found from the VA, 
with those, found from the imaginary-time method. 
For $S= 0$, the profiles, obtained by the two methods, are very close to each other in a wide range
of $N$. For $S= 0$, we choose $N = 1000$ in order to demostrate a flat-top shape of QDs.
We take $N =60, 200$ and 510 for vortices with $S =1, 2$ and 3,
respectively. We mention that the norm values chosen for $S > 0$ correspond to the stability thresholds 
(see below). Figure~\ref{fig:prof} shows that the proposed approach gives a good 
agreement for stationary solutions.
A difference between the predicted and exact profiles increases for larger $S$. We find 
that Eq.~(\ref{ansatz}) gives a reasonable approximation of stationary solutions with vorticity 
up to $S =5$.

\begin{figure}[htbp]
\centerline{ \includegraphics[width=4.2cm]{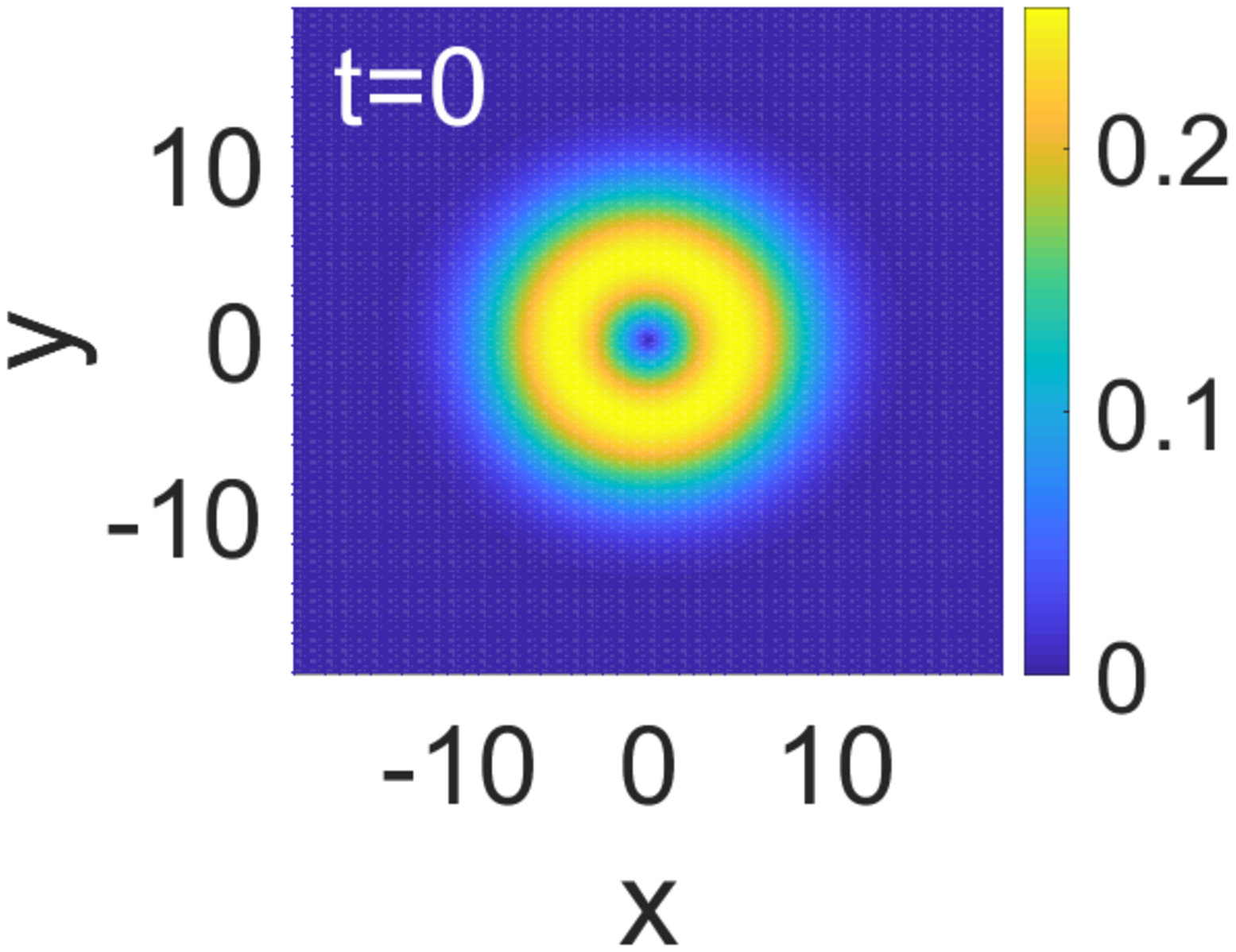}  \includegraphics[width=4.2cm]{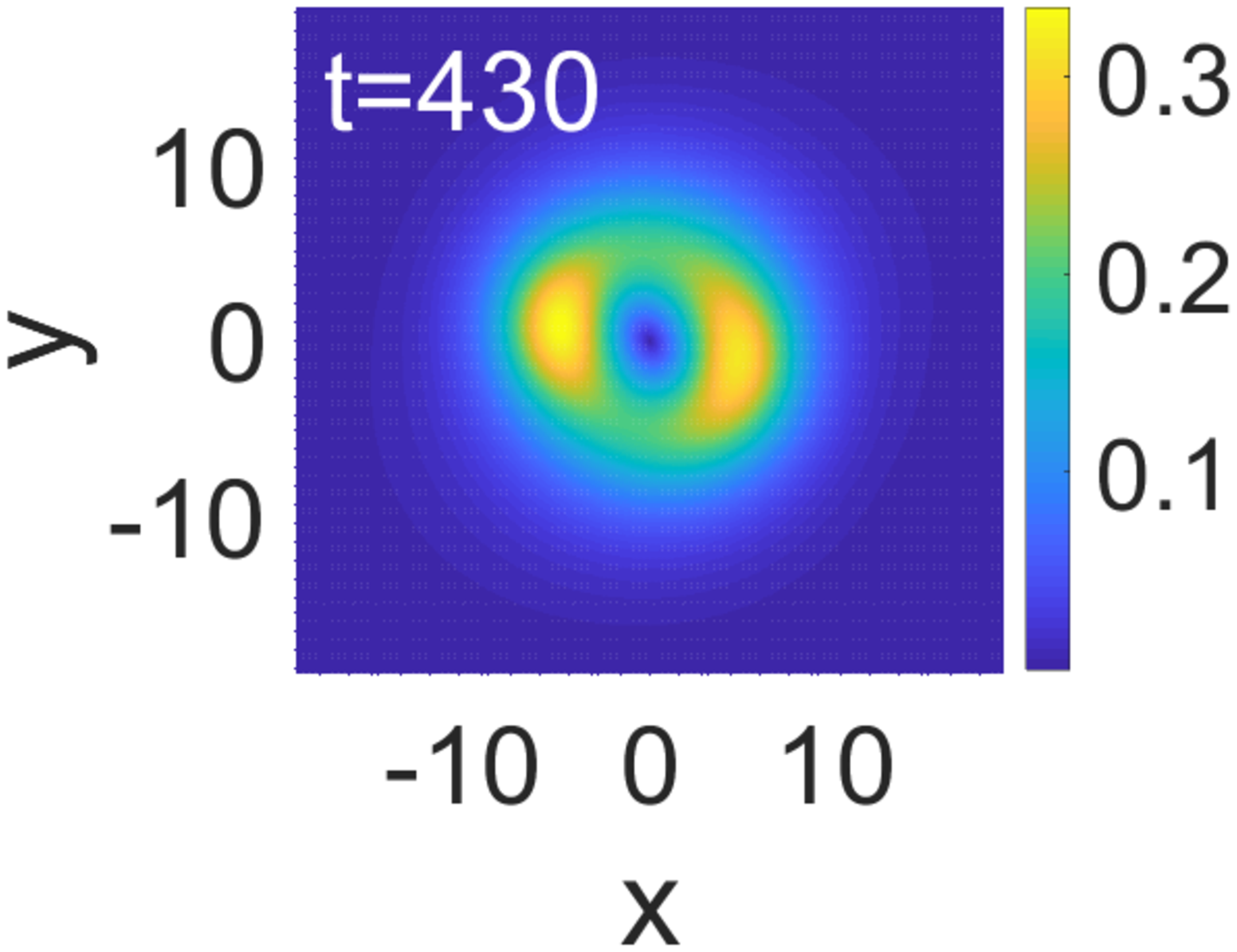} }
\centerline{ \includegraphics[width=4.2cm]{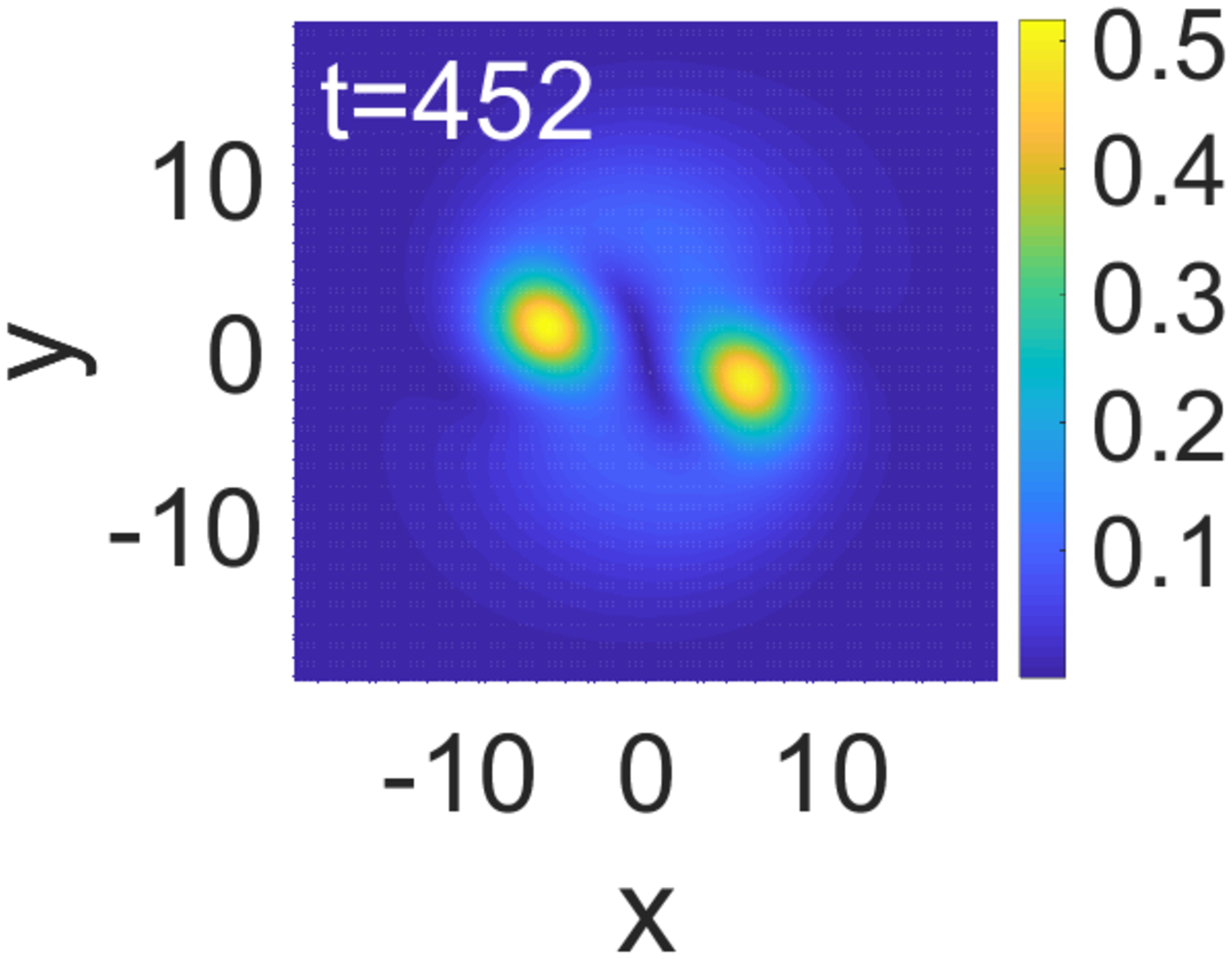} \includegraphics[width=4.2cm]{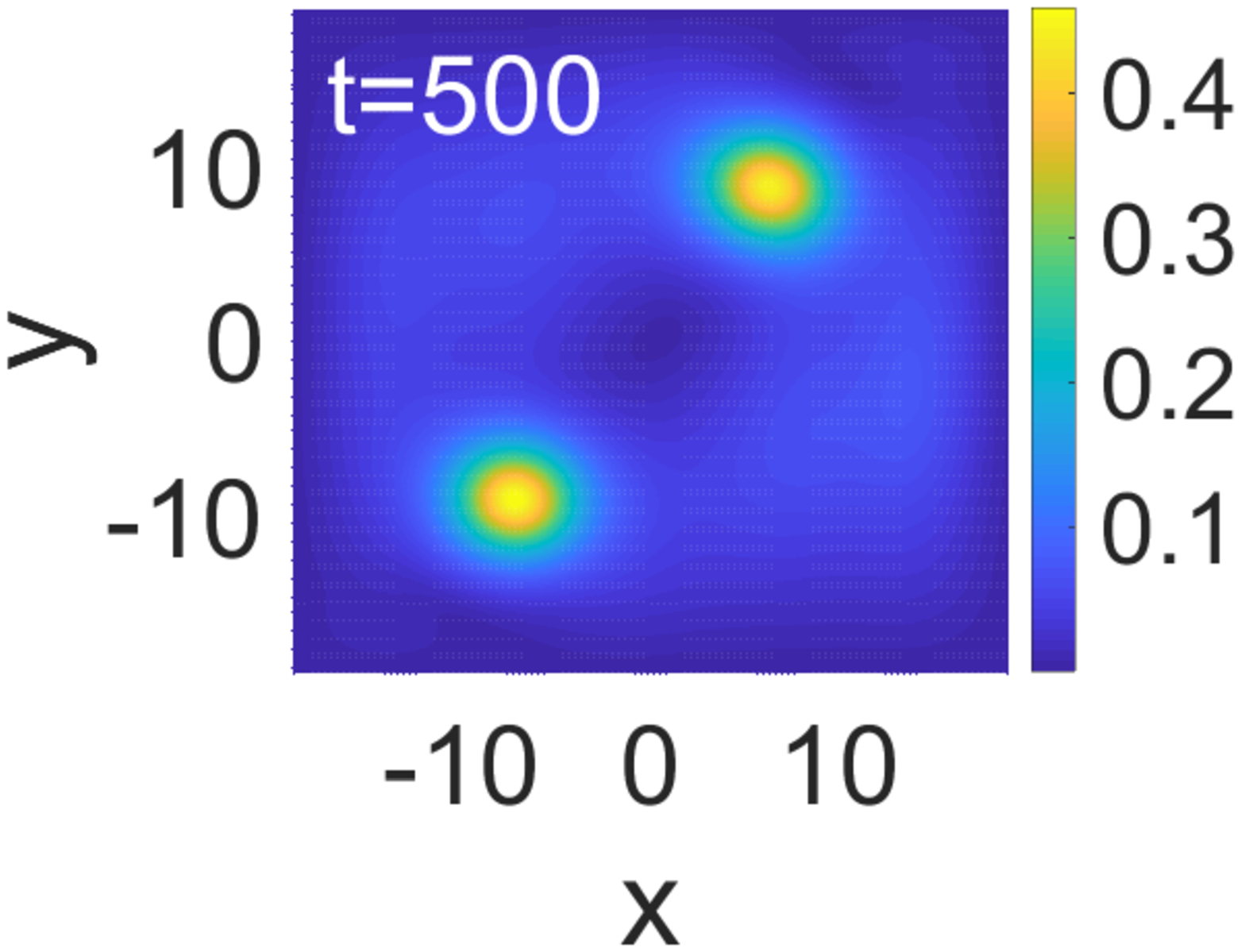} }
\caption{(color online) The evolution of $|\Psi|$ 
of the unstable vortex QD for $S = 1$, $N=10$, $m=1.232$, $A=0.082$
($\bar{A}=0.253$), and $w=5.051$. }
\label{fig:dyn}
\end{figure}

   According to the Vakhitov-Kolokolov criterion~\cite{Vakh73}, a negative value of the
derivative, $d\mu / dN <0$, means the stability of QDs. Then, as it follows from the inset
in Fig.~\ref{fig:u_mu}, QDs with any $S$ are stable. However this result is valid only
for $S = 0$, because the Vakhitov-Kolokolov criterion gives a necessary condition, and it
accounts only for the stability along $r$ direction, not on $\theta$. Numerical simulations, using the 
SSF method, of Eq.~(\ref{gpe}) up to $t=10^4$ show that QDs with $S = 0$ are stable, while 
vortices with $S > 0$ can be unstable~\cite{Li18}.
A typical scenario of an unstable vortex is shown in Fig.~\ref{fig:dyn}.
As an initial condition of Eq.~(\ref{gpe}), we use Eq.~(\ref{ansatz}) with stationary parameters,
found from the VA. The vortex parameters oscillate slightly, however the vortex form
is preserved basically until $t \sim 300$. Then the unstable vortex splits into two QDs with $S = 0$ each.
These QDs have both the radial and tangential velocities, however the total momentum
is constant (zero). An unstable vortex with vorticity $S$ split typically into $S+1$
fragments~\cite{Li18}.

  For a given $0 < S \leq 5$, there is a threshold value $N_{\mathrm{th}}$, below which the vortex 
is unstable~\cite{Li18}. We confirm this result numerically. For different values of N, we integrate 
Eq.~(\ref{gpe}) with initial conditions in a form of Eq.~(\ref{ansatz}) and stationary 
parameters, found from the VA. We identify a vortex as stable, if there no splitting up to $t \sim 10^4$.
We find that for $S = 1, 2$, and $3$, the stability thresholds are $N_{\mathrm{th}}= 60, 200$, and $510$, 
respectively. This coincide with results of Ref.~\cite{Li18}, where it is found also that no stable QDs
exist for $S > 5$ and $N \lesssim 10^4$. In Figs.~\ref{fig:u_mu} and~\ref{fig:pars}, parameters
of unstable QDs are shown by dashed lines.

   In Ref.~\cite{Li18}, it was stated also that there is another threshold
$N_{\mathrm{min}}(S)$ for $0 <S \leq 5$ below which vortices do not exist. In contrast to this work,
we find that vortices exist for $N < N_{\mathrm{min}}$ as well, though they are unstable. Initial
conditions, found from Eq.~(\ref{ansatz}) with the stationary parameters for
$0 < S \leq 5$, remain almost unchanged up to $t \sim 300$ even for small $N$.
This indicates that the initial condition, found from the VA, is close to a stationary (unstable) 
state. In other words, the initial dynamics of vortices for $N < N_{\mathrm{min}}$ is similar to
the dynamics of unstable vortices for $N_{\mathrm{min}} < N < N_{\mathrm{th}}$.
For larger $t$, the vortex splitting occurs, as described above. We mention that the imaginary time method, 
used in Ref.~\cite{Li18}, requires an additional tuning of parameters for 
finding {\em unstable} stationary solutions. 

\begin{figure}[htbp]
  \centerline{ \includegraphics[width=5.5cm]{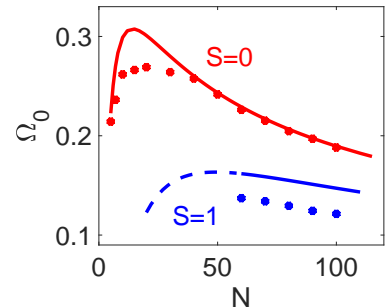}}
\caption{(color online) The angular frequency $\Omega_0$ of small oscillations of $w(t)$ for
$S = 0$ and $1$. Lines are found from Eq.~(\ref{freq}), while points are found from numerical
simulations. A dashed line corresponds to unstable vortices.
}
\label{fig:per}
\end{figure}

  The VA gives also good predictions of the dynamical properties of QDs.
Figure~\ref{fig:per} illustrates the dependence of the angular frequency $\Omega_0$ of small internal
oscillations on $N$. We use Eq.~(\ref{ansatz}) with slightly (1--5\%) perturbed stationary
parameters as an initial condition in numerical simulations of Eq.~(\ref{gpe}). We observe
long-lived (up to $t \sim 1000$) oscillations of the QD shape.
We measure the period $\tau$ from the dependence of $w(t)$ after $4\textrm{--}5$ oscillations as an 
average over 5--10 periods, and the angular frequency is found as $\Omega_0 = 2\pi/\tau$.
Figure~\ref{fig:per} demonstrates a good agreement for $S = 0$, and a
reasonable agreement for $S = 1$.

\subsection{Periodic variation of $\delta(t)$}
\label{sec:param}

   In order to analyze deeper the relevance of the VA, we study the QD dynamics under the
action a periodic modulation of parameter $\delta(t)$
\begin{equation}
  \delta(t) = \delta_{0} [1 + \epsilon \sin(\omega_{m} t)],
\label{delt}
\end{equation}
where $\epsilon$ and $\omega_{m}$ are the amplitude and the angular frequency of
modulations, respectively. Such modulations can be created by a periodic variation of the
external magnetic field via the Feschbach resonance.

\begin{figure}[htbp]
   \centerline{ \includegraphics[width=8.6cm]{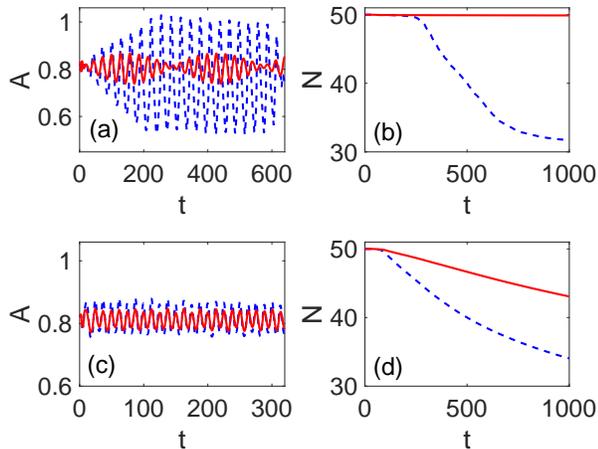}}
\caption{(color online) The dependencies of amplitude $A$ [(a) and (c)] and norm $N$
[(b) and (d)] on time, found from numerical
simulations of Eq.~(\ref{gpe}) for $\delta_{0} = -1$, $S=0$, and $N=50$.
(a)  and (b) $\omega_{m} = 0.21$, $\epsilon = 0.5 \epsilon_1$
(solid lines),  and $\epsilon = \epsilon_1$  (dashed lines);
(c)  and (d) $\omega_{m} = 0.5$, $\epsilon = 0.5 \epsilon_2$
(solid lines),  and $\epsilon = \epsilon_2$  (dashed lines), where
$\epsilon_1 = 0.187$ and $\epsilon_2 = 0.252$.
}
\label{fig:osc}
\end{figure}

   The periodic variation of $\delta(t)$ induces oscillations of the QD parameters,
such as $A(t)$ and $w(t)$. We analyze the dynamics of the fundamental QD ($S = 0$) for $N =50$.
For given $\epsilon$, the resonance frequency $\omega_r$ can be found from the dependence
of the amplitude difference $\Delta A \equiv A_{\mathrm{max}} -  A_{\mathrm{min}}$ on
frequency $\omega_m$, where $A_{\mathrm{max}}$ ($A_{\mathrm{min}}$) is the larges (lowest)
value of $A(t)$ on time. At $\omega_m = \omega_r$, the amplitude difference has a peak.
We find, by solving numerically Eq.~(\ref{gpe}), that for small $\epsilon$, frequency $\omega_r$ is close
to the eigenfrequency $\Omega_0$. Frequency $\omega_r$ changes with an increase of $\epsilon$
due to nonlinearity.

\begin{figure}[htbp]
   \centerline{ \includegraphics[width=6cm]{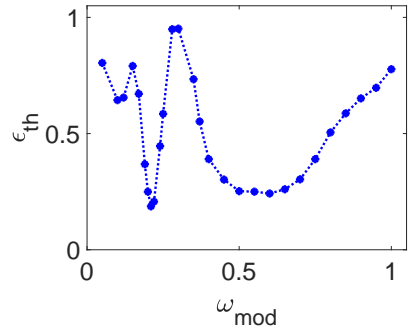}}
\caption{
Threshold $\epsilon_{\mathrm{th}}$ as a function of $\omega_{m}$, found
from numerical simulations of Eq.~(\ref{gpe})
for $\delta_{0} = -1$, $S=0$, and $N=50$.
}
\label{fig:thresh}
\end{figure}

   We consider firstly the QD dynamics for the driving frequency close to $\Omega_0(N=50) = 0.243$.
In Fig.~\ref{fig:osc}a, we plot the dependence of $A$ on $t$ for
$\omega_{m} = 0.21$, $\epsilon = 0.5 \epsilon_1$ and $\epsilon = \epsilon_1$,
where $\epsilon_1 = 0.187$. A choice of such $\omega_m$ and $\epsilon_1$  will be explained below,
see also Fig.~\ref{fig:thresh}. When $\omega_{m} = 0.21$ and $\epsilon = 0.5 \epsilon_1$, we
observe a beating of the QD amplitude, see a solid line in Fig.~\ref{fig:osc}a. This
beating is typical for forced oscillations, and it is consistent  with the VA approach.
Namely, a change of  $\delta(t)$ results in a deformation of the potential $U(w)$, see
Eq.~(\ref{pot}), and a periodic variation of the minimum point. This deformation of $U(w)$ gives
rise to oscillations of the QD parameters. Since $N$ almost does not change, see
a solid line in Fig.~\ref{fig:osc}b, the dynamics is adiabatic.

  An increase of $\epsilon$ results usually in oscillations with larger $\Delta A$, as in
Fig.~\ref{fig:osc}a for $\omega_{m} = 0.21$ and $\epsilon = \epsilon_1$ (a dashed line).
For these parameters, an increase of the amplitude envelope on time has initially a linear slope that
indicates a resonance in oscillations. Oscillations of $A(t)$ saturate due to nonlinear effects.
Large oscillations of the QD amplitude result in a strong excitation of the internal mode.
The QD starts to emit particles that is observed as round outgoing waves of the BEC
density. A dashed line in Fig.~\ref{fig:osc}b shows that the emission of waves begins at $t \approx 200$,
and after that N decreases due to the absorbing boundary conditions. Large oscillations results
in a QD splitting into two smaller droplets that move in opposite directions. For $\epsilon
= \epsilon_1$, the splitting occurs at $t \sim 650$. This type of the dynamics is typical
for frequencies  $\omega_{m} \sim \Omega_0$, or for large $\epsilon$.
Value $\epsilon_1$ is taken such that $N$ loses 20 \% at $t = 500$.  We use later a
corresponding condition to analyze the dynamics for different $\omega_m$. We mention that
splitting of solitons was found also for the nonlinear Schr\"{o}dinger model with the varying
coefficient of the 2BI~\cite{Saka04}, see also Ref.~\cite{Abdu03}.

   Now we consider the dynamics out of the resonance for $\omega_{m} = 0.5$,
$\epsilon = 0.5\epsilon_2$ and $\epsilon = \epsilon_2$, where $\epsilon_2 = 0.252$,
see Figs.~\ref{fig:osc}c and~\ref{fig:osc}d. After a development of oscillations, see
Fig.~\ref{fig:osc}c, the QD
emits particles in a form of linear matter waves. This fact follows from a decrease of $N$
starting at $t \sim 100$ in Fig.~\ref{fig:osc}d. The larger values of $\epsilon$ induce larger
oscillations, and therefore a faster decay of $N$, see dashed lines for $\epsilon= \epsilon_2$ 
in Figs.~\ref{fig:osc}c and~\ref{fig:osc}d. Here, $\epsilon_2$ is taken, using the same condition 
as for $\epsilon_1$. 

  For $\epsilon = \epsilon_2$, function $N(t)$ at large $t$ changes very little, and it has some signs 
of saturation, approaching value $N= 22.3$ at $t = 10^4$. For small $\epsilon$, it takes much longer 
time to see saturation of $N(t)$, cf. the solid and dashed lines in Fig.~\ref{fig:osc}d. Thus, for 
small $\epsilon$, we observe that after an emission of an appreciable amount of particles, $N(t)$ 
changes slowly at large $t$. This means that a response of a QD to periodic modulations of 
$\delta$ out of the resonance becomes adiabatic at large $t$. For sufficiently large $\epsilon$, 
$\epsilon> \epsilon_2$, the asymptotic value of $N(t)$ can be less than 2--5\% of its initial value. 
This corresponds to a substantial evaporation of a QD under the action of periodic modulations.
 
   An analysis of numerical results for different values of $\omega_{m}$ and $\epsilon$
reveals four main types of the QD dynamics. In the first type, the QD oscillates
adiabatically, with negligible emission of waves. This type occurs for small $\epsilon$.
The VA, presented in Sect.~\ref{sec:model}, is valid for this adiabatic regime. In all of
the rest types of the dynamics, the QD emits linear waves after a steady period. This
emission of waves corresponds to evaporation of the QD. In the second type of the dynamics,
a QD transfers at large $t$ to the adiabatic regime with smaller $N$.
A decrease of $N$ on time changes the parameters of the
stationary state, see Fig.~\ref{fig:pars}, and therefore the response to the driving field.
In the third type, a QD decays completely via emission of linear waves. For some parameters,
this decay takes large time $t >\sim 10^4$. In the fourth type, a QD splits into smaller
moving droplets. Since $N$ decreases on time for the last three types of the dynamics, the VA is
not applicable for these cases.
The first type and the second type are similar, they differ only by the
amount of particles, remaining in the QD. In order to separate the adiabatic
regimes (the first and the second types) from the decay and the splitting (the third and the
fourth types), we introduce the following criterion. For given $\omega_{m}$, we find such
$\epsilon_{\mathrm{th}}$ that $N$ decreases at $t = 500$ to 80\% from the initial value,
and the decrease continues (no saturation), at least, up to $t = 1000$.

  The dependence of $\epsilon_{\mathrm{th}}$ on $\omega_{m}$ is presented in
Fig.~\ref{fig:thresh}. We can say that $\epsilon_{\mathrm{th}}$ gives an approximate
threshold between different regimes. The adiabatic regimes exist for sure far below the threshold, 
while the QD splitting or a complete QD decay exist much above this value.
There is a narrow minimum near $\omega_{m} = 0.21$,
and a wide deep for $\omega_{m} = [0.4, 0.7]$. The minimum near  $\omega_{m} = 0.21$
corresponds to resonance oscillations, where the splitting may occur even for small $\epsilon$. 
We see that the VA gives a proper estimate of the resonance frequency. Curve $\epsilon_{\mathrm{th}}$ 
in interval $\omega_{m} = [0.4, 0.7]$ separates the regime of small and large changes of $N$.

\section{Conclusions}
\label{sec:conc}

   We have shown, that similarly to the 1D case~\cite{Otaj19}, the super-Gaussian function
is a good approximation for 2D quantum droplets. The dynamical equations for the QD parameters
have been derived using the VA. Analytical equations for parameters of stationary droplets
with $S = 0$ and  $S = 1$ have been obtained. The density of stationary QDs tends to a
constant for large $N$, while the width grows gradually. Such a property characterizes a QD
as a cluster of incompressible liquid. Our analysis confirms results of work~\cite{Li18} that QDs with
$S = 0$ are stable, while vortices with $0 < S \leq 5$ are unstable, when $N <
N_{\mathrm{th}}(S)$.

   It was also demonstrated that the VA gives a proper description of the QD dynamics. The
frequency $\Omega_0$ of small internal oscillations of QDs has been obtained. Values of
$\Omega_0$ gives good (reasonable) approximation for $S = 0$  ($S = 1$). It has been found
that modulations of $\delta$ with frequency close $\Omega_0$ induces resonance
oscillations of the QD
parameters. We have identified different regimes of the dynamics, including adiabatic
oscillations, the decay of QDs, and the QD splitting, for different parameters of
modulations.

   The case $\delta >0$ has been analyzed as well. The VA predicts the existence of
stationary vortices, but they are unstable. Vortices either spread dispersively or
collapse, depending on a value of the initial width (amplitude).

\section*{Acknowledgements}
   This work was supported by grant FA-F2-004 of the Ministry of Innovative Development
of the Republic of Uzbekistan.



\begin{thebibliography}{99}

\bibitem{Peth08} C. J. Pethick and H. Smith, Bose-Einstein Condensation in
Dilute Gases (Cambridge University Press, Cambridge, 2008).

\bibitem{Kart19}    Ya. V. Kartashov, G. E. Astrakharchik, B. A. Malomed, and L. Torner,
Frontiers in multidimensional self-trapping of nonlinear fields and matter,
Nature Rev. Phys. {\bf 1}, 185 (2019).

\bibitem{Petr15} D. S. Petrov, Quantum mechanical stabilization of a
collapsing Bose-Bose mixture, Phys. Rev. Lett. {\bf 115}, 155302 (2015).

\bibitem{Petr16} D. S. Petrov and G. E. Astrakharchik, Ultradilute low-dimensional liquids,
Phys. Rev. Lett. {\bf 117}, 100401 (2016).

\bibitem{LHY} T. D. Lee, K. Huang, and C. N. Yang, Eigenvalues and eigenfunctions
of a Bose system of hard spheres and its low-temperature properties,
Phys. Rev. {\bf 106}, 1135 (1957).

\bibitem{Cabr18} C. R. Cabrera, L. Tanzi, J. Sanz, B. Naylor, P. Thomas, P.Cheiney, and
L. Tarruell, Quantum liquid droplets in a mixture of Bose-Einstein condensates,
Science {\bf 359}, 301 (2018).

\bibitem{Seme18} G. Semeghini, G. Ferioli, L. Masi, C. Mazzinghi, L. Wolswijk, F. Minardi,
M. Modugno, G. Modugno, M. Inguscio, and M. Fattori, Self-bound quantum droplets
in atomic mixtures, Phys. Rev. Lett. {\bf 120}, 235301 (2018).

\bibitem{Ferr16} I. Ferrier-Barbut, H. Kadau, M. Schmitt, M. Wenzel, and T.
Pfau, Observation of quantum droplets in a strongly dipolar
Bose gas, Phys. Rev. Lett. {\bf 116}, 215301 (2016).

\bibitem{Li18} Y. Li, Z. Chen, Z. Luo, C. Huang, H. Tan, W. Pang, and B. A. Malomed,
Two-dimensional vortex quantum droplets, Phys. Rev. A {\bf 98}, 063602 (2018).

\bibitem{Kart19a}  Ya. V. Kartashov, B. A. Malomed, and L. Torner, Metastability of quantum
droplet clusters, Phys. Rev. Lett. {\bf 122}, 193902 (2019).

\bibitem{Garm64} R. Y. Chiao, E.  Garmire,  and  C. H. Townes, Self-trapping of optical beams,
Phys. Rev. Lett. {\bf 13}, 479 (1964).

\bibitem{Otaj19} Sh. R. Otajonov, E. N. Tsoy, F. Kh. Abdullaev, Stationary and dynamical
properties of one-dimensional quantum droplets, Phys. Lett. A {\bf 383}, 125980  (2019).

\bibitem{Vakh73} N. G. Vakhitov and A. A. Kolokolov, Stationary solutions of
the wave equation in a medium with nonlinearity saturation, Radiophys.
Quantum Electron. {\bf 16}, 783 (1973).

\bibitem{Saka04} H. Sakaguchi and B. A. Malomed, Resonant nonlinearity management
for nonlinear Schr\"{o}dinger solitons, Phys. Rev. E {\bf 70}, 066613 (2004).

\bibitem{Abdu03} F. Kh. Abdullaev, E. N. Tsoy, B. A. Malomed, R. A. Kraenkel,
Array of Bose-Einstein condensates under time-periodic Feshbach-resonance
management, Phys. Rev. A {\bf 68}, 053606 (2003).

\end{thebibliography}
\end{document}